%% file: iChannel_Feedback_ArXiv.tex
\documentclass[11pt, onecolumn, draftclsnofoot]{IEEEtran}
\usepackage{fancyhdr}
\usepackage{amsmath,epsfig}
\usepackage{threeparttable}
\usepackage{epsf,epsfig}
\usepackage{amsthm}
\usepackage{amsmath}
\usepackage{amssymb}
\usepackage{amsfonts}
\usepackage{algorithmic}
\usepackage{algorithm}
\usepackage[noadjust]{cite}
\usepackage{dsfont}
\usepackage{subfigure}
\usepackage{color}

\input{header}

\begin{document}

\title{\huge \setlength{\baselineskip}{30pt} Cooperative Precoding with Limited Feedback for MIMO Interference  Channels }
\author{\large \setlength{\baselineskip}{15pt}Kaibin Huang and Rui Zhang  \thanks{ K. Huang is  with the School of Electrical and Electronic Engineering,   Yonsei University, S.  Korea. 
R. Zhang is with the Department of Electrical and Computer Engineering, National University of Singapore, and the Institute for Infocomm Research,
A*STAR, Singapore. Email: huangkb@ieee.org, elezhang@nus.edu.sg. 
This work was supported in part by National Research Foundation of Korea under the grant 2011-8-0740 and the National University of Singapore under the grant R-263-000-589-133. 
}\vspace{-40pt}}

\maketitle

\begin{abstract} 
Multi-antenna precoding effectively  mitigates the interference in  wireless networks. However, the resultant performance gains can be significantly compromised  in practice if the precoder design fails to account for the inaccuracy in the  channel state information (CSI) feedback. This paper addresses this issue by considering  finite-rate CSI feedback from receivers to their interfering transmitters in the two-user multiple-input-multiple-output (MIMO) interference channel, called \emph{cooperative feedback}, and  proposing a systematic method for designing  transceivers comprising linear precoders and equalizers. Specifically, each precoder/equalizer  is decomposed into inner and outer components for nulling the cross-link interference and achieving   array gain, respectively. The inner precoders/equalizers  are further optimized  to suppress the residual interference resulting from  finite-rate cooperative feedback. 
Furthermore, the residual interference is regulated by additional scalar cooperative feedback signals that are designed to control transmission power using different criteria including fixed  interference margin and maximum  sum throughput. Finally, the  required number of cooperative precoder feedback bits is  derived for limiting  the throughput loss  due to precoder quantization.   
\end{abstract}

\section{introduction}
In wireless networks, multi-antennas can be employed to  effectively mitigate  interference between coexisting links by precoding.  This paper presents a new precoding design for  the two-user  multiple-input-multiple-output (MIMO) interference channel  based on finite-rate channel-state-information (CSI) exchange between users, called \emph{cooperative feedback}. Specifically, precoders are designed to suppress interference to interfered receivers based on their quantized CSI feedback, and the residual interference is regulated by additional cooperative feedback of  power control signals.

\subsection{Prior Work}

Recently,  progress has been made on analyzing  the capacity   of the multi-antenna  interference channel. In particular,  interference alignment techniques have been proposed for achieving the channel capacity for high signal-to-noise ratios (SNRs)   \cite{CadJafar:InterfAlignment:2007}. Such techniques, however, are impractical  due to their complexity, requirement of perfect global CSI, and their sub-optimality for finite SNRs. This prompts the development of linear precoding algorithms for practical decentralized wireless networks. For the time-division duplexing (TDD)  multiple-input-single-output (MISO) interference channel, it is proposed in \cite{ZakhourGesbert:DistMutlicellMISOPrecodingLayerVirtualSINR, DahYu:CoordBeamformMulticell:2010} 
that  the forward-link beamformers can be adapted distributively based on reverse-link  signal-to-interference-and-noise ratios (SINRs).   
Targeting the two-user MIMO interference channel, linear transceivers are designed  in \cite{ChaeHeath:InterfAwareCoordBeamformTwoCell} under the constraint of one data stream per user and using different criteria including zero-forcing and minimum-mean-squared-error. In \cite{ZhangCui:CoopIMMISOBeamform},  the achievable rate region for the MISO interference channel is analyzed based on the  interference-temperature principle in cognitive radio, yielding  
a  message passing  algorithm for enabling distributive  beamforming. Assuming perfect transmit  CSI, the above prior work does not address the issue of finite-rate CSI feedback though it is widely used in  precoding implementation. Neglecting  feedback CSI errors in precoder designs can result in over-optimistic network performance. 

For MIMO precoding  systems, the substantiality of  CSI feedback overhead has motivated extensive   research on efficient CSI-quantization  algorithms, forming a field called \emph{limited feedback} \cite{Love:OverviewLimitFbWirelssComm:2008}. Various limited feedback algorithms  have been proposed based on different principles such as  line packing
\cite{LovHeaETAL:GrasBeamMultMult:Oct:03} and Lloyd's algorithm \cite{Lau:MIMOBlockFadingFbLinkCapConst:04}, which were  applied to design   specific MIMO systems including  beamforming
\cite{LovHeaETAL:GrasBeamMultMult:Oct:03} and 
precoded spatial multiplexing
\cite{LoveHeath:LimitedFeedbackPrecodSpatialMultiplex:05}. Recent limited feedback research has focused on  MIMO downlink systems, where multiuser CSI feedback  supports space-division multiple access \cite{Gesbert:ShiftMIMOParadigm:2007}.  It has been found that the number  of feedback bits  per user has to increase with the transmit SNR so as to bound the throughput loss caused by feedback quantization \cite{Jindal:MIMOBroadcastFiniteRateFeedback:06}.
Furthermore, such a loss  can be reduced  by exploiting \emph{multiuser diversity} \cite{SharifHassibi:CapMIMOBroadcastPartSideInfo:Feb:05, Huang:OrthBeamSDMALimtFb:07}. Designing limited feedback algorithms  for the interference channel is more challenging  due to the decentralized network architecture and the growth of total feedback CSI. 
Cooperative feedback algorithms are proposed in \cite{HuangZhang:CoopFeedbackCognitiveRadio} for 
a two-user  cognitive-radio network, where the secondary   transmitter adjusts its beamformer to suppress interference to  the primary   receiver that cooperates by feedback to the secondary transmitter.  This design is tailored for a MISO cognitive radio network and thus unsuitable for the general MIMO interference channel, which motivates this  work. 

\subsection{Contributions}

The precoder design that maximizes the sum throughput of the MIMO interference channel is a non-convex optimization problem and remains open \cite{Gesbert:MultiCellMIMOCooperativeNetworks:2010}. In practice, sub-optimal linear procoders are commonly used for their simplicity, which are designed assuming perfect transmit CSI and based on various criteria including interference suppression by zero-forcing or minimum transmission power for given received SINRs \cite{Gesbert:MultiCellMIMOCooperativeNetworks:2010}. However, existing designs fail to  exploit the interference-channel realizations for suppressing residual interference due to quantized cooperative feedback. In this work, we consider the two-user MIMO interference channel with limited feedback and propose the decomposed  precoder design that makes it possible for precoding to simultaneously  regulate  residual interference due to precoder-feedback  errors and enhance   received signal power. For the purpose of exposition, we consider two coexisting MIMO links where each link employs $L$  transmit and $K$ receive antennas to support  multiple data streams. Linear precoding is applied at each transmitter and  enabled by quantized cooperative feedback.  Channels are modeled as  i.i.d. Rayleigh block fading.

The main contributions of this work are summarized as follows:
\begin{enumerate}
\item A systematic method is proposed for jointly designing the linear precoders and equalizers  under the zero-forcing criterion, which  decouples the links in the event of perfect  feedback. To be specific, precoders and equalizers are decomposed into inner and outer components that are  designed to suppress residual interference caused by feedback errors and enhance array gain, respectively. 

\item Additional  scalar cooperative feedback, called \emph{interference power control} (IPC) feedback, is proposed for controlling  transmission power so as to regulate  residual interference. The IPC feedback algorithms are designed using  different criteria  including fixed interference margin and  maximum sum throughput.
   
\item Consider  cooperative feedback of inner precoders of the size $L\times N_p$ with $N_p\leq L$. Under a constraint on  the throughput loss caused by precoder quantization, the required  number of feedback bits is   shown to scale linearly with $N_p(L-N_p)$ and logarithmically with the transmit SNR as it increases. 

\end{enumerate}

Despite both addressing cooperative feedback for the two-user interference channel, this work differs from   \cite{HuangZhang:CoopFeedbackCognitiveRadio} in the following aspects. The current system comprises two MIMO links whereas that in \cite{HuangZhang:CoopFeedbackCognitiveRadio} consists of a SISO and a MISO links. Correspondingly, this paper and \cite{HuangZhang:CoopFeedbackCognitiveRadio}  concern  precoding and transmit  beamforming, respectively. Furthermore, 
this work does not consider cognitive radio as in \cite{HuangZhang:CoopFeedbackCognitiveRadio}, leading to different design principles. In particular, the current  system requires CSI exchange between two links while that in \cite{HuangZhang:CoopFeedbackCognitiveRadio} involves only one-way cooperative feedback from the primary receiver to the secondary transmitter.

\subsection{Organization}
The remainder of this paper is organized as follows. The system model is discussed  in Section~\ref{Section:System}. The transceiver design and IPC feedback algorithms are presented in  Section~\ref{Section:Orth:Algo} and \ref{Section:IPC}, respectively. The feedback requirements are analyzed in Section~\ref{Section:FbRequire}. Simulation results  are presented in Section~\ref{Section:Simulation} followed by concluding remarks.

{\bf Notation:} Capitalized and small boldface letters denote matrices and vectors, respectively.  The superscript $\dagger$ represents the  Hermitian-transpose matrix operation. The operators $[\bX]_{k}$ and $[\bX]_{mn}$ give the $k$-th column and the $(m, n)$-th element  of a matrix $\bX$, respectively. Moreover,  $[\bX]_{m:n}$   with $n\geq m$ represents  a matrix formed by columns $m$ to  $n$ of the matrix $\bX$. 
%Let $\preceq$, $\prec$,  $\succeq$ and $\succ$ represent element-wise inequalities between two real vectors.  
The  operator $(\cdot)^+$ is defined as $(a)^+ = \max(a, 0)$ for $a\in \mathds{R}$.

\section{System Model}\label{Section:System}
\begin{figure}
\begin{center}
\includegraphics[width=10cm]{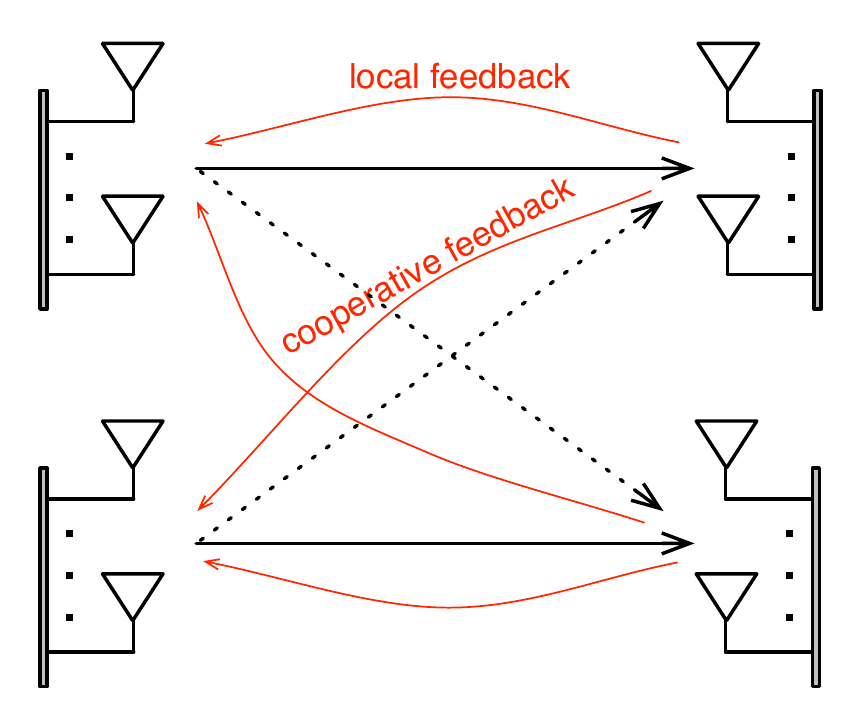}
\caption{The MIMO interference channel with data-link (local) and cooperative feedback   }
\label{Fig:Sys}
\end{center}
\end{figure}

We consider two interfering wireless links as illustrated in Fig.~\ref{Fig:Sys}.  Each transmitter and receiver employ   $L$  and $K$ antennas, respectively,  to  suppress interference  as well as supporting spatial multiplexing. These operations require CSI feedback from receivers to their interfering and intended transmitters, called \emph{cooperative feedback} and \emph{data-link feedback}, respectively. 
 We assume perfect   CSI estimation and  data-link feedback, allowing the current design to focus on suppressing interference 
 caused by  cooperative feedback quantization.\footnote{The errors in data-link feedback decrease  received SNRs, which  can be compensated by increasing transmission power. }  All channels are assumed to follow independent block fading.  The channel coefficients are   i.i.d. circularly symmetric complex Gaussian  random variables with zero mean and unit variance, denoted as $\mathcal{CN}(0,1)$.   Let $\bH_{mn}$ be a $K\times L$ i.i.d. $\mathcal{CN}(0,1)$ matrix representing  fading in the channel from transmitter $n$ to receiver $m$. Then the interference channels are modeled as  $\{\nu\bH_{mn}\}$  with  $m\neq n$ and the data channels as  $\{\bH_{mm}\}$. The factor $\nu < 1$ quantifies the path-loss difference  between the data and interference links. 
 
Each link supports  $M\leq \min(L, K)$ spatial data streams by linear precoding and equalization. To regulate residual interference caused by precoder feedback errors, the total transmission power of each transmitter is controlled  by cooperative IPC feedback. For simplicity, the scalar IPC feedback is assumed to be perfect since it requires much less overhead than the precoder feedback.    Each transmitter uses identical transmission  power for  all spatial streams,  represented by  $P_n$  for   transmitter $n$ with $n=1, 2$,  and its maximum is denoted as $P_{\max}$. Assume that all  additive white noise samples are i.i.d.  $\mathcal{CN}(0, 1)$ random variables. Let $\bG_m$ and $\bF_m$ denote the linear equalizer used by receiver $m$ and the linear precoder applied at transmitter $m$, respectively, which are jointly designed for separating the  spatial  data streams of user $m$ with $m=1, 2$. 
The  receive  SINR   at  receiver $m$ for the $\ell$th stream can be written as 
\begin{equation}
\SINR_m^{[\ell]}  = \frac{ P_m\|[\bG_m]_\ell^\dagger \bH_{mm}[\bF_m]_\ell\|^2}{1 + \nu P_n \|[\bG_m]_\ell^\dagger \bH_{mn}\bF_n\|^2 }, \quad m\neq n. 
\label{Eq:SINR}
\end{equation}
The  performance metric is the sum throughput defined as 
\begin{equation} \label{Eq:ErgCap}
\bar{C} = \sum_{m=1}^2\sum_{\ell =1}^M \E\l[\log_2\l(1+\SINR_m^{[\ell]} \r)\r].  
\end{equation}
%Next, consider the scenario where the coding rates for all data streams are fixed  at $\log_2(1+\theta)$ where $\theta$ is the receive SINR threshold for correct decoding. We define an outage event as one that the SINR of at least one data stream is smaller than $\theta$. It follows that the  outage probability  is given by
%\begin{equation}
%\Pout =\Pr\l(\min_{m=1,2}\min_{1\leq \ell\leq M}\SINR_m^{[\ell]}< \theta\r).  \label{Eq:Pout}
%\end{equation}

\section{Transceiver Design}\label{Section:Orth:Algo}
In this section, we propose a decomposition approach for designing the transceivers (linear precoders and equalizers).   Using this approach, the precoder $\bF_n$ is decomposed  into  an \emph{inner precoder} $\bF_n^{\mathsf{i}}$ and an \emph{outer precoder} $\bF_n^{\mathsf{o}}$.\footnote{The decomposed precoder designs have been proposed in the literature for other systems such as cellular downlink \cite{Chae:BDVPMuMIMO:2008}. }
 Specifically, $\bF_n  = \bF_n^{\mathsf{i}}\bF_n^{\mathsf{o}}$ where  $\bF^{\mathsf{i}}_n$  and $\bF_n^{\mathsf{o}}$ are $L\times N_p$ and $N_p\times M$ matrices, respectively, with $N_p$ being  no smaller than the number of data streams $M$ and $N_p \leq L$.  Similarly, we decompose the equalizer $\bG_m$ as $\bG_m  = \bG_m^{\mathsf{i}}\bG_m^{\mathsf{o}}$ where  $\bG_m^{\mathsf{i}}$ is an $K\times N_e$ \emph{inner equalizer} and  $\bG_m^{\mathsf{o}}$  an $N_e \times M$ \emph{outer equalizer} with  $M \leq N_e\leq K$. For simplicity, inner/outer precoders and equalizers are constrained to have orthonormal columns. The inner and outer transceivers are designed to suppress cross-link interference  and achieve array gain, respectively. In the following sub-sections, the transceivers, namely the inner/outer precoders and equalizers, are first designed   assuming perfect cooperative feedback and then modified to mitigate residual interference caused by feedback quantization. 

\subsection{Transceiver Design for Perfect Cooperative Feedback}

\subsubsection{Inner Transceiver Design}
A  pair of inner precoder and equalizer $(\bG_m^{\mathsf{i}}, \bF^{\mathsf{i}}_n)$ with $m\neq n$ are jointly designed under the following zero-forcing criterion:
\begin{equation}\label{Eq:OrthConst:b}
(\bG_m^{\mathsf{i}})^\dagger \bH_{mn} \bF_n^{\mathsf{i}} = \mathbf{0},\quad  m\neq n. 
\end{equation}
The constraint aims at decoupling  the links and requires that $N_e + N_p  \leq \max(L, K)$.  Under the constraint in \eqref{Eq:OrthConst:b}, $(\bG_m^{\mathsf{i}}, \bF^{\mathsf{i}}_n)$ with $m\neq n$ are  designed by decomposing $\bH_{mn}$ using the  singular value decomposition (SVD) as 
\begin{equation}
\bH_{mn} =\bV_{mn} \boldsymbol{\Sigma}_{mn}\bU_{mn}^\dagger
\label{Eq:H:SVD:a}
\end{equation}
where the unitary matrices  $\bV_{mn}$ and $\bU_{mn}$ consist of the left and right singular vectors of $\bH_{mn}$ as columns, respectively, and  $\boldsymbol{\Sigma}_{mn}$ is a diagonal matrix with diagonal elements $\{\sqrt{\lambda_{mn}^{[\ell]}}\}$ arranged in the descending order, namely $\lambda_{mn}^{[1]}\geq \lambda_{mn}^{[2]}\cdots \geq \lambda_{mn}^{[\min(L, K)]}$. Note that $\boldsymbol{\Sigma}_{mn}$ is a tall matrix if $K \geq L$ and a fat one if $K < L$. Definite the index sets $\mathcal{A}\subset \{1, 2, \cdots, K\}$ and $\mathcal{B}\subset \{1, 2, \cdots, L\}$ such that $|\mathcal{A}|=N_e$ and  $|\mathcal{B}| = N_p$. 
 The constraint in \eqref{Eq:OrthConst:b} can be satisfied if $\mathcal{A}\cap \mathcal{B}=\emptyset$ and the inner precoder and equalizer are chosen as 
\begin{equation}
\bG_m^{\mathsf{i}} = [\bV_{mn}]_{\mathcal{A}} \ \textrm{and}\ \bF_n^{\mathsf{i}}=[\bU_{mn}]_{ \mathcal{B}}. \label{Eq:Inner:Design:a}
\end{equation}
Consider the case of $K \geq N_e + N_p$. Each receiver has sufficiently  many antennas  for canceling cross-link interference and thus cooperative feedback is unnecessary. Specifically, given an arbitrary fixed precoder $\bF_n^{\mathsf{i}}$, the equalizer $\bG_m^{\mathsf{i}}$ chosen as in \eqref{Eq:Inner:Design:a} ensures that the zero-forcing criterion in \eqref{Eq:OrthConst:b} is satisfied. Next, consider the case of $K < N_e + N_p$. For this case,  the receivers have insufficient degrees of freedom (DoF) for canceling cross-link interference and link decoupling  relies  on inner precoding that is feasible given that $N_e + N_p \leq L$. Therefore, 
with $K<L$,\footnote{This condition usually holds for cellular downlink where a base station has more antennas than a mobile terminal.}  
cooperative feedback is in general required and the specific design of inner transceiver for quantized cooperative  feedback  is discussed in the sequel.

 \subsubsection{Outer Transceiver Design} \label{Section:Orth:Algo:Perfect}

 Given $(\bG_m^{\mathsf{i}}, \bF_m^{\mathsf{i}})$, the outer pair $(\bG_m^{\mathsf{o}}, \bF_m^{\mathsf{o}})$ are jointly designed based on the SVD of the $N_e\times N_p$  effective channel $\bH_{mm}^{\mathsf{o}}=\l(\bG_m^{\mathsf{i}}\r)^\dagger \bH_{mm}\bF_m^{\mathsf{i}}$ after inner precoding and equalization: 
\begin{equation}
\bH_{mm}^{\mathsf{o}} = \bV_{mm}^{\mathsf{o}} \boldsymbol{\Sigma}_{mn}^{\mathsf{o}}(\bU_{mm}^{\mathsf{o}})^\dagger   \nn
\end{equation}
where the singular values $\sqrt{\lambda_{mm}^{[1]}},  \sqrt{\lambda_{mm}^{[2]}}, \cdots, \sqrt{\lambda_{mm}^{[\min(N_e, N_p)]}}$ follow the descending order. 
Note that the elements of $\bH_{mm}^{\mathsf{o}}$ are i.i.d. $\mathcal{CN}(0, 1)$ random variables and their distributions are independent of $(\bG_m^{\mathsf{i}}, \bF_m^{\mathsf{i}})$ since $\bH_{mm}$ is isotropic.  Transmitting  data through the strongest eigenmodes of $\bH_{mm}^{\mathsf{o}}$ enhances the received SNR. This can be realized by choosing   $\bG_m^{\mathsf{o}}$ and  $\bF_m^{\mathsf{o}}$ as 
\begin{equation}
\bG_m^{\mathsf{o}} = [\bV_{mm}^{\mathsf{o}}]_{1:M}\quad \textrm{and}\quad \bF_m^{\mathsf{o}} = [\bU_{mm}^{\mathsf{o}}]_{1:M}.\nn
\end{equation} 
With perfect data-link  feedback, the above  joint design of precoders and equalizers converts each data link into $M$ decoupled  spatial channels. As a result, the receive SNR of the $\ell$-th data stream transmitted from  transmitter $m$ to receiver $m$ is given by 
\begin{equation}
\SNR^{[\ell]}_m = P_m\lambda^{[\ell]}_{mm}, \qquad \ell = 1, 2, \cdots, M. \label{Eq:SNR}
\end{equation}
Using the maximum transmission power, the sum capacity $\tilde{C}$ can be written as 
\begin{equation}\label{Eq:SumRate:PerfCSI}
\tilde{C} = \sum_{m=1}^2\sum_{\ell=1}^M \E\l[\log_2(1 + P_{\max}\lambda^{[\ell]}_{mm})\r].  
\end{equation}
%and the outage probability as 
%\begin{equation}\label{Eq:Pout:PerfCSI}
%\aPout = \Pr \l(P_{\max}\min\l(\lambda_{11}^{[M]}, \lambda_{22}^{[M]}\r) < \theta \r). 
%\end{equation}

Last, it is worth mentioning  that with $L$, $K$ and $M$ fixed, maximizing  $N_p$ and $N_e$  enhances the array gain of both links and hence is preferred if perfect CSI is available at the transmitters. 
 However,  for  the case of quantized cooperative feedback, small $N_p$ and $N_e$ allow more DoF to  be used for suppressing 
residual interference due to precoder-quantization errors as discussed in the next section. 

\subsection{Transceiver Design for Quantized  Cooperative Feedback}
\label{Section:Orth:Algo:Quant} 
Consider the case of $K < N_e + N_p$ and mitigating cross-link interference relies on inner precoding with quantized feedback. As mentioned earlier,  cooperative feedback is unnecessary if $K \geq N_e + N_p$. 

\subsubsection{Inner Transceiver Design} 
In this section, the joint design of inner precoders and equalizers in \eqref{Eq:Inner:Design:a}  is modified to  suppress the residual interference caused by precoder feedback errors. 

First, given the inner equalizer $\bG_m^{\mathsf{i}}$ in \eqref{Eq:Inner:Design:a}, the inner precoder $\bF_n^{\mathsf{i}}$ in \eqref{Eq:Inner:Design:a} is particularized under the criterion of minimizing residual interference power. 
Recall that the precoding at  transmitter $n$ is  enabled by quantized cooperative feedback of $\bF_n^{\mathsf{i}}$ from receiver $m$ with $m\neq n$. Let $\hat{\bF}_n^{\mathsf{i}}$ denote the quantized version of  $\bF_n^{\mathsf{i}}$  that is also an orthonormal matrix. Define the  resultant quantization error $\epsilon_n$  as \cite{RavJindal:LimFbDiagonalMIMOBC:2008}
 \begin{equation}\label{Eq:QuantErr}
\epsilon_n = 1 - \frac{\|(\bF_n^{\mathsf{i}})^\dagger \hat{\bF}_n^{\mathsf{i}}\|^2_{\mathsf{F}}}{N_p}, \quad n = 1, 2
\end{equation}
where  $0\leq  \epsilon_n \leq 1$. The  error $\epsilon_n$ is zero in the case of perfect cooperative feedback, namely $\bF_n^{\mathsf{i}} = \hat{\bF}_n^{\mathsf{i}}$. A nonzero error results in the violation of the zero-forcing criterion in \eqref{Eq:OrthConst:b}
\begin{equation}\label{Eq:OrthConst:a}
\l(\bG_m^{\mathsf{i}}\r)^\dagger\bH_{mn}\hat{\bF}_n^{\mathsf{i}} \neq 0,\quad m\neq n. 
\end{equation}
Given that the inner equalizer designed for perfect feedback is applied, 
the residual interference at the output of the equalizer at receiver $m$ has the power 
\begin{equation}
I_{m} = \nu P_n | (\bG_m^{\mathsf{o}})^\dagger (\bG_m^{\mathsf{i}})^\dagger \bH_{mn} \hat{\bF}_n^{\mathsf{i}}\bF_n^{\mathsf{o}}|^2_{\mathsf{F}},\quad  m \neq n.  
\label{Eq:Interf}
\end{equation}
It is difficult to directly optimize  $\hat{\bF}_n^{\mathsf{i}}$ for minimizing $I_m$. Alternatively, $\hat{\bF}_n^{\mathsf{i}}$ can be designed for minimizing an upper bound on $I_m$ obtained as follows. By rearranging eigenvalues and eigenvectors, the SVD of $\bH_{mn}$ in \eqref{Eq:H:SVD:a} for $L > K$ can be rewritten as 
\begin{equation}
\bH_{mn} =\l[\begin{matrix}\bG_{m}^{\mathsf{i}}& \bB_m\end{matrix}\r] \l[ \begin{matrix}   \sqrt{\boldsymbol{\Sigma}_{mn}^{(a)} } &   
 \boldsymbol{0} &  \boldsymbol{0}\\
  \boldsymbol{0} & \sqrt{\boldsymbol{\Sigma}_{mn}^{(a)} } &   
 \boldsymbol{0}  
\end{matrix} \r] \l[\begin{matrix}\bC_{n} & \bF_n^{\mathsf{i}}\end{matrix}\r]^\dagger
\label{Eq:H:SVD:b}
\end{equation}
where the diagonal matrices $\boldsymbol{\Sigma}_{mn}^{(a)}$ and $\boldsymbol{\Sigma}_{mn}^{(b)}$ have the diagonal elements $\{\sqrt{\lambda_{mn}^{[\ell]}}\mid \ell \in\mathcal{A}\}$ and $\{\sqrt{\lambda_{mn}^{[\ell]}}\mid 1\leq \ell \leq K,  \ell \notin\mathcal{A}\}$, respectively. 
It follows from \eqref{Eq:Interf}  and \eqref{Eq:H:SVD:b} that 
\begin{align}
I_{m} &= \nu P_n \l\| (\bG_m^{\mathsf{o}})^\dagger \sqrt{\boldsymbol{\Sigma}_{mn}^{(a)}} \bC^\dagger_n \hat{\bF}_n^{\mathsf{i}}  \bF_n^{\mathsf{o}} \r\|^2_{\mathsf{F}}\nn\\
&\leq \nu P_n \l\| (\bG_m^{\mathsf{o}})^\dagger \sqrt{\boldsymbol{\Sigma}_{mn}^{(a)}} \bC^\dagger_n \hat{\bF}_n^{\mathsf{i}}   \bU_{nn}^{\mathsf{o}} \r\|^2_{\mathsf{F}}\nn\\
&= \nu P_n \l\| (\bG_m^{\mathsf{o}})^\dagger \sqrt{\boldsymbol{\Sigma}_{mn}^{(a)}} \bC^\dagger_n \hat{\bF}_n^{\mathsf{i}} \r\|^2_{\mathsf{F}}\label{Eq:I:UB:a}\\
&= \nu P_n \sum_{\ell=1}^{M}\sum_{k=1}^{N_p}\l| [\bG_m^{\mathsf{o}}]_{\ell}^\dagger \sqrt{\boldsymbol{\Sigma}_{mn}^{(a)}} \bC^\dagger_n [\hat{\bF}_n^{\mathsf{i}}]_k \r|^2\nn\\
&\leq \nu P_n \sum_{\ell=1}^{M}\sum_{k=1}^{N_p}\| [\bG_m^{\mathsf{o}}]_{\ell}\|^2\l\|\sqrt{\boldsymbol{\Sigma}_{mn}^{(a)}} \bC^\dagger_n [\hat{\bF}_n^{\mathsf{i}}]_k \r\|^2\label{Eq:I:UB:b}\\
&=  M \nu P_n\l\|\sqrt{\boldsymbol{\Sigma}_{mn}^{(a)}} \bC^\dagger_n \hat{\bF}_n^{\mathsf{i}}\r \|^2_{\mathsf{F}}\label{Eq:I:UB:c}\\
&\leq M \nu P_n\l\|\bC^\dagger_n \hat{\bF}_n^{\mathsf{i}}\r \|^2_{\mathsf{F}}\max_{\ell \in \mathcal{A}}\lambda_{mn}^{[\ell]}\label{Eq:Interf:Ub:a}
\end{align}
where \eqref{Eq:I:UB:a} holds since the columns of $\bU_{nn}^{\mathsf{o}}$ form a basis of the space $\mathds{C}^L$, \eqref{Eq:I:UB:b} applies Schwarz's inequality and \eqref{Eq:I:UB:c} follows from that the columns of $\bG_m^{\mathsf{o}}$ have unit norms. Next, the precoder quantization error $\epsilon_n$  in \eqref{Eq:QuantErr} can be written as
\begin{eqnarray}
\epsilon_n &=& \frac{1}{N_p}\sum_{\ell=1}^{N_p}\l(1 - \|[\hat{\bF}_n^{\mathsf{i}}]^\dagger_\ell\bF_n^{\mathsf{i}} \|^2_{\mathsf{F}}\r)\nn\\
&=& \frac{1}{N_p}\sum_{\ell=1}^{N_p}\|[\hat{\bF}_n^{\mathsf{i}}]^\dagger_\ell\bC_n \|^2_{\mathsf{F}}\label{Eq:QErr:a}\\
&=& \frac{1}{N_p}\|\bC_n^\dagger \hat{\bF}_n^{\mathsf{i}}   \|^2_{\mathsf{F}}\label{Eq:QErr}
\end{eqnarray}
where \eqref{Eq:QErr:a} holds since $[\bF_n^{\mathsf{i}}, \bC_n]$ forms a basis of the space $\mathds{C}^L$.  Substituting \eqref{Eq:QErr} into \eqref{Eq:Interf:Ub:a} gives 
\begin{equation}
I_m \leq \nu M N_p P_n \epsilon_n \max_{\ell \in \mathcal{A}}\lambda_{mn}^{[\ell]}\label{Eq:Interf:Ub:b}. 
\end{equation}
Minimizing the right-hand side of  \eqref{Eq:Interf:Ub:b} gives that the columns of $\bG_m^{\mathsf{i}}$ should be left eigenvectors of $\bH_{mn}$ corresponding to the $N_e$  smallest  singular values. Therefore, the inner transceiver design in \eqref{Eq:Inner:Design:a} for perfect CSI is particularized as 
\begin{equation}
\bG_m^{\mathsf{i}} =[\bV_{mn}]_{\mathcal{A}}\ \text{with} \ \mathcal{A} = \{K-N_e+1, \cdots, K\} \quad \text{and}\quad \bF_n^{\mathsf{i}} =[\bU_{mn}]_{\mathcal{B}} \ \text{with}\ \mathcal{B}\cap \mathcal{A} = \emptyset \label{Eq:Inner:Design}
\end{equation}
and $\hat{\bF}_n^{\mathsf{i}}$ is obtained by quantizing $\bF_n^{\mathsf{i}}$ such that the quantization error $\epsilon_n$ is minimized \cite{RavJindal:LimFbDiagonalMIMOBC:2008}. 
Then \eqref{Eq:Interf:Ub:b}  can be simplified as 
\begin{equation}
I_{m} \leq  \nu M N_p P_n\lambda_{mn}^{[K-N_e+1]}\epsilon_n, \quad m\neq n. \label{Eq:Interf:Ub:c}
\end{equation}

Next, if $K > N_e$,   besides $N_e$ DoF required for  inner equalization, a receiver has $(K-  N_e)$ extra DoF that can be used to suppress residual interference. This can be realized at receiver $m$ by redesigning the inner equalizer $\bG_m^{\mathsf{i}}$ with the resultant design denoted as  $\hat{\bG}_m^{\mathsf{i}}$. To this end, the matrix $\bH_{mn}\hat{\bF}_n^{\mathsf{i}}$ is  decomposed by SVD as 
\begin{equation}
\bH_{mn}\hat{\bF}_n^{\mathsf{i}} =\hat{\bV}_{mn} \hat{\boldsymbol{\Sigma}}_{mn}\hat{\bU}_{mn}^\dagger 
\nn
\end{equation}
where the singular values along the diagonal of  $\hat{\boldsymbol{\Sigma}}_{mn}$ are denoted as $\{\sqrt{\hat{\lambda}_{mn}^{[\ell]}}\mid \ell \leq \min(K, N_p)\}$ and  arranged in the descending order. 
The inner equalizer $\hat{\bG}_m^{\mathsf{i}}$ that minimizes the residual interference should be chosen to comprise the left eigenvectors of $\bH_{mn}\hat{\bF}_n^{\mathsf{i}}$ that correspond to the smallest singular values and hence  $\hat{\bG}_m^{\mathsf{i}} = [\hat{\bV}_{mn}]_{(K-N_e+1):K}$. Another interpretation of this design is that the inner equalizer $\hat{\bG}_{m}^{\mathsf{i}}$ is directed towards the null space of  $\hat{\bF}_n^{\mathsf{i}}$. The resultant  residual interference power after inner precoding can be upper bounded as 
\begin{align} 
\tilde{I}_m &= \nu P_n\|\hat{\bG}_{m}^{\mathsf{i}} \bH_{mn}\hat{\bF}_n^{\mathsf{i}}\bF_n^{\mathsf{o}}\|^2_{\mathsf{F}}\nn\\
&\leq \nu P_n\|\hat{\bG}_{m}^{\mathsf{i}}  \bH_{mn}\hat{\bF}_n^{\mathsf{i}}\|^2_{\mathsf{F}}\nn\\
&=\nu P_n\sum_{m=K-N_e+1}^{N_p}\sqrt{\hat{\lambda}_{mn}^{[\ell]}}. \label{Eq:I:Ub:QCSI}
\end{align}
  Last, if $K = N_e$, the design of $\bG_m^{\mathsf{i}}$ remains unchanged and   $\hat{\bG}_m^{\mathsf{i}}= \bG_m^{\mathsf{i}}$.

\subsubsection{Outer Transceiver Design}
Let $\hat{\bG}_m^{\mathsf{o}}$ and $\hat{\bF}_m^{\mathsf{o}}$ denote the outer equalizer and  precoder for the case of quantized cooperative feedback. The outer transceiver ($\hat{\bG}_m^{\mathsf{o}}$ and $\hat{\bF}_m^{\mathsf{o}})$ is designed similarly as its perfect-feedback counterpart in Section~\ref{Section:Orth:Algo:Perfect}. Decompose the effective channel matrix $\l(\hat{\bG}_m^{\mathsf{i}}\r)^\dagger \bH_{mm}\hat{\bF}_m^{\mathsf{i}}$ after inner precoding/equalization as
\begin{equation}
\l(\hat{\bG}_m^{\mathsf{i}}\r)^\dagger \bH_{mm}\hat{\bF}_m^{\mathsf{i}} = \hat{\bV}_{mm}^{\mathsf{o}}\hat{\boldsymbol{\Sigma}}_{mm}^{\mathsf{o}} (\hat{\bU}_{mm}^{\mathsf{o}})^\dagger  \label{Eq:H:SVD:11}
\end{equation}
where the diagonal matrix  $\hat{\boldsymbol{\Sigma}}_{mm}^{\mathsf{o}}$ contains singular values $\{\sqrt{\hat{\lambda}_{mm}^{\ell}}\}$ arranged  in the descending order. 
To maximize the received SNRs, the outer equalizer and precoder are chosen as 
\begin{equation}
\hat{\bG}_m^{\mathsf{o}} = [\hat{\bV}_{mm}^{\mathsf{o}}]_{1:M}\quad \textrm{and}\quad \hat{\bF}_m^{\mathsf{o}} = \hat{\bU}_{mm}^{\mathsf{o}}.\nn
\end{equation} 
The corresponding sum throughput follows  from \eqref{Eq:ErgCap} as
\begin{equation} \label{Eq:ErgCap:a}
\bar{C} = \sum_{m=1}^2\sum_{\ell =1}^M \E\l[\log_2\l(1+\frac{ P_m\hat{\lambda}_{mm}^{\ell}}{1 + \nu P_n \|[\bG_m]_\ell^\dagger \bH_{mn}\bF_n\|^2 } \r)\r]
\end{equation}
where $\bG_m = \hat{\bG}_m^{\mathsf{i}}\hat{\bG}_m^{\mathsf{o}}$ and $\bF_n = \hat{\bF}_n^{\mathsf{i}}\hat{\bF}_n^{\mathsf{o}}$. 

\subsection{Discussion}
In this section, we discuss the robustness of the proposed quantized feedback precoder by comparison with  a conventional design  without cooperative feedback. To be specific, the baseline design is the well-known single-user transceiver design that provides  no cooperative  feedback, where interference is treated as noise and all spatial DoF are applied to maximize array gain \cite{PaulrajBook}. Let the data-channel matrix $\bH_{mm}$ be decomposed by SVD as $\bH_{mm} = \bV_{mm}\boldsymbol{\Sigma}_{mm}\bU_{mm}^\dagger$ with $m \in\{1, 2\}$.   The precoder $\acute{\bF}_m$ and receiver $\acute{\bG}_m$ for the baseline case as given below transmit data through the $M$ strongest eigenmodes of $\bH_{mm}$ \cite{PaulrajBook}
\begin{equation}\label{Eq:Algorithm:b}
\acute{\bF}_m = [\bV_{mm}]_{1:M}\quad  \text{and} \quad \acute{\bG}_m = [\bU_{mm}]_{1:M}. 
\end{equation}
By using  the maximum transmission power, the  corresponding sum throughput is given as 
\begin{eqnarray}
\acute{C} &=& 2\E\l[\sum_{\ell=1}^M \log_2\l(1 + \frac{P_{\max}|\l[\boldsymbol{\Sigma}_{11}\r]_{\ell, \ell} |^2}{1 + P_{\max}\nu |[\acute{\bG}_1]_1^\dagger \bH_{12}\acute{\bF}_2|^2 } \r)\r]\nn\\
&\leq& 2\E\l[\sum_{\ell=1}^M \log_2\l(1 + \frac{|\l[\boldsymbol{\Sigma}_{11}\r]_{\ell, \ell} |^2}{\nu |[\acute{\bG}_1]_\ell^\dagger \bH_{12}\acute{\bF}_2|^2 } \r)\r]. \label{Eq:SumCap:Cmp}
\end{eqnarray}
Similarly as \eqref{Eq:Interf:Ub:c}, it can be proved that the interference power for the $\ell$-th data stream, $I_m^{[\ell]}$,  received at receiver $m$ is upper funded as 
\begin{equation}
I_{m}^{[\ell]} \leq  \nu N_p P_n\lambda_{mn}^{[K-N_e+1]}\epsilon_n, \quad m\neq n. \label{Eq:Interf:Ub:d}
\end{equation}
 From  \eqref{Eq:ErgCap:a} and \eqref{Eq:Interf:Ub:d} and also using the maximum transmission  power,   the sum throughput $\bar{C}$ for the proposed design can be lower bounded as 
\begin{equation}\label{Eq:SumCap:LB:Cmp}
\bar{C} \geq 2\E\l[\sum_{\ell=1}^M \log_2\l(1 + \frac{P_{\max}\lambda_{11}^{[\ell]}}{1 + N_p P_{\max}\nu \lambda_{12}^{[K-N_e+1]}\epsilon_2} \r)\r]. 
\end{equation}
By comparing \eqref{Eq:SumCap:Cmp} and \eqref{Eq:SumCap:LB:Cmp}, it can be observed that $\acute{C}$ is bounded as  $P_{\max}$ increases but  $\bar{C}$ can grow with increasing $P_{\max}$ if the quantization errors $\{\epsilon_m\}$ are regulated by adjusting the number of feedback bits based on $P_{\max}$  (see Section~\ref{Section:FbRequire} for details).

\section{Interference Power Control  Feedback} \label{Section:IPC}

In this section, we consider the case of $K = N_e$ where receivers have no extra DoF for suppressing residual interference. An alternative solution is to adjust transmission power for increasing the sum throughput. Two IPC feedback algorithms for implementing power control are discussed in the following sub-sections. 

\subsection{Fixed Interference Margin}\label{Section:FixMargin}
Receiver  $m$ sends the IPC signal, denoted as $\eta_n$, to transmitter/interferer $n$ for controlling its transmission power as 
\begin{equation}
P_n  = \min(\eta_n, P_{\max}), \quad n = 1, 2. \label{Eq:TxPwr:IPC}
\end{equation}
The scalar $\eta_n$ is  designed to prevent the per-stream interference power at receiver $m$  from exceeding a fixed margin $\tau$ with $\tau > 0$, namely $I_{m}^{[\ell]} \leq \tau$ for all $0\leq \ell \leq M$. A sufficient condition for satisfying this constraint is to bound   the right hand side of \eqref{Eq:Interf:Ub:d}  by   $\tau$. It follows that 
\begin{equation}
\eta_n = \frac{\tau}{N_p\nu\lambda_{mn}^{[K-N_e+1]}\epsilon_n}, \quad m\neq n. \label{Eq:IPC}
\end{equation}
Given $\tau$, a lower bound $A_{\mathsf{IM}}$ on the  sum throughput  $\bar{C}$, called \emph{achievable throughput},   is  obtained from \eqref{Eq:ErgCap:a} as 
\begin{equation}
A_{\mathsf{IM}} =  \sum_{m=1}^2\sum_{\ell =1}^M \log_2\l(1 + \frac{\min(\eta_m, P_{\max}) \lambda_{mm}^{[\ell]}}{1+\tau}\r).  \label{Eq:SumCap:LB}
\end{equation}

It is infeasible to derive  the optimal value of $\tau$ for maximizing $A_{\mathsf{IM}}$ in \eqref{Eq:SumCap:LB}. However, for $P_{\max}$ being either large or small, simple insight into choosing $\tau$ can be derived   as follows. The residual interference power decreases continuously with reducing $P_{\max}$. Intuitively, $\tau$ should be kept small for small $P_{\max}$. For large $P_{\max}$, the choice of $\tau$ is less intuitive  since large $\tau$ lifts the constraints on the transmission power  but  causes stronger interference and vice versa. We show below that large $\tau$ is preferred for large $P_{\max}$. This requires the result in  \cite[Theorem~1]{OrdeonezPalomar:HighSNRPerformMIMO:2007} paraphrased  as follows.
\begin{lemma}[\cite{OrdeonezPalomar:HighSNRPerformMIMO:2007}]\label{Lem:Wishart} Let $\bH$ denote a $Q_1\times Q_2$ matrix of i.i.d. $\mathcal{CN}(0, 1)$ elements with $Q_1 \geq Q_2$. The cumulative distribution  function of the $k$-th eigenvalue $\phi_k$ of the Wishart matrix $\bH^\dagger\bH$ can be expanded as 
\begin{equation}
 \Pr(\phi_k < x ) =  a_kx^{d_k} + o(x^{d_k}),\quad k = 1, \cdots, Q_2
\end{equation}
where $d_k = (Q_1 - k +1)(Q_2-k+1) $ and $a_k = \frac{U^{-1} |\bA(k)||\bB(k)|}{d_k}$ with $U = \prod_{m=1}^{Q_2} (Q_1-m)!(Q_2 - m)!$. The matrix $\bA(k)$ is defined for $k\neq 1$ as 
\begin{equation}
[\bA(k)]_{mn} = (Q_1-Q_2+m+n + 2(Q_2-k))!, \quad m, n = 1, \cdots, (k-1)
\end{equation}
and $\bA(1) = \bI$, $\bB(k)$ is defined for $k \neq Q_2$ as 
\begin{equation}
[\bB(k)]_{mn} = \frac{2}{[(Q_1-Q_2 + m+n)^2 - 1](Q_1-Q_2 + m+n)}, \quad m, n = 1, \cdots, (Q_2-k)
\end{equation}
and $\bB(Q_2) = \bI$.
\end{lemma}
\noindent To simplify notation, we re-denote $(\phi_k, a_k)$ for $Q_1= N_e$ and $Q_2 = N_p$ as $(\acute{\lambda}_k, \acute{a}_k)$ and those for $Q_1= Q_2=L$ as $(\check{\lambda}_k, \check{a}_k)$. Using the above result, we obtain the following lemma that is  proved in the appendix.  
\begin{lemma} \label{Lemma:SumCap:IM} { Given finite-rate cooperative feedback and for large $P_{\max}$},  the achievable throughput is  
\begin{equation}\label{Eq:Thput:IM:LargeP}
A_{\mathsf{IM}}  =
 2\sum_{\ell = 1}^M  \E\l[\log_2\l(1\!+ \!\frac{\tau\acute{\lambda}_\ell}{(1+\tau)N_p\nu\check{\lambda}_{K-N_e+1}\epsilon_1}\r)\r] + o\l(1\r).
\end{equation}
\end{lemma}
\noindent It can be observed from the above result that  the first order term of $A_{\mathsf{IM}}$ attains its maximum as $\tau\rightarrow\infty$. However, this term  is finite even for asymptotically large  $P_{\max}$ and $\tau$, which is the inherent effect of residual interference.  
%\begin{lemma} \label{Lem:Pout:IM}For large $P_{\max}$,  the outage probability  is upper bounded as  
%\begin{equation}
%\Pout  \leq2\Pr\l( \frac{\tau}{1+\tau}\times \frac{\acute{\lambda}_{M}}{M\nu\check\lambda_{L-N+1}\epsilon_1}< \theta \r) + o(1).\nn  \label{Eq:PoutIM:LargeP}
%\end{equation}
%\end{lemma}
%\noindent Similar remarks on Lemma~\ref{Lemma:SumCap:IM} apply to Lemma~\ref{Lem:Pout:IM}. 

\subsection{Maximum Achievable Throughput}\label{Section:IPC:Cap} \label{Section:ErgCap}
In this section, an iterative  IPC algorithm is designed for increasing the sum throughput  $\bar{C}$ in 
\eqref{Eq:ErgCap}. 
Since  $\bar{C}$ is a non-concave function of transmission power, directly maximizing $\bar{C}$ does not yield a simple IPC algorithm. Thus, we resort  to maximizing a lower bound $A_{\mathsf{ST}}$ (achievable throughput) on $\bar{C}$ instead,  obtained  from \eqref{Eq:Interf:Ub:c}   and \eqref{Eq:ErgCap:a} as 
$A_{\mathsf{ST}} =  \E\l[A\r]$ with 
\begin{equation}\label{Eq:Cap:Lb}\begin{aligned}
&A &=& \ \sum_{\ell=1}^M\left[  \log_2\l(1+ \frac{P_1 \lambda_{11}^{[\ell]}}{1 + P_2 N_p\nu \lambda_{12}^{[K-N_e+1]} \epsilon_2}\r)+\right.\\
&&&\left. \log_2\l(1+\frac{ P_2 \lambda_{22}^{[\ell]}}{1 + P_1N_p\nu \lambda_{21}^{[K-N_e+1]}\epsilon_1}\r)\r]. 
\end{aligned}
\end{equation}
The corresponding optimal transmission power pair is given as 
\begin{equation}
(P_1^\star, P_2^\star) = \arg\max_{P_1, P_2\in [0, P_{\max}]} A(P_1, P_2). 
\end{equation}
The objective function  $A$ remains non-concave and its maximum has no known closed-form for $M > 1$. Note that for $M=1$, it has been shown that the optimal  transmission-power pair belongs to the set $\{(0, P_{\max}), (P_{\max}, 0), (P_{\max}, P_{\max})\}$ \cite{GjenGesbert:BinaryPowerConrolInterferenceLinks:2008}. 
For the current case of $M > 1$, inspired by the message passing algorithm in \cite{ZhangCui:CoopIMMISOBeamform},   a  sub-optimal search for $(P_1^\star, P_2^\star)$  can be derived using  the fact that 
\[
\frac{\partial A(P_1^\star, P_2^\star)}{\partial P_m}   = 0,\quad  \forall\ m = 1, 2. 
\]
To this end, the slopes of $A$ are obtained using \eqref{Eq:Cap:Lb} as 
\begin{equation}
\frac{\partial A(P_1, P_2)}{\partial P_m}  = \mu_{m} + \psi_m - \rho_{m}\label{Eq:CapSlope}
\end{equation}
where 
\begin{eqnarray}
\mu_{m} &=& \log_2e\sum\nolimits_{\ell=1}^M \frac{\lambda_{mm}^{[\ell]}}{1 + N_p\nu\lambda_{mn}^{[K-N_e+1]}\epsilon_nP_n + \lambda_{mm}^{[\ell]}P_m}\nn\\
\psi_{m} &=& \log_2e\sum\nolimits_{\ell=1}^M \frac{N_p\nu\lambda_{nm}^{[K-N_e+1]}\epsilon_m}{1 + N_p\nu\lambda_{nm}^{[K-N_e+1]}\epsilon_mP_m + \lambda_{nn}^{[\ell]}P_n}\nn\\
\rho_{m} &=&   \frac{\log_2e N_p^2\nu\lambda_{nm}^{[K-N_e+1]}\epsilon_m}{1+N_p\nu\lambda_{nm}^{[K-N_e+1]}\epsilon_mP_m}.   \nn
\end{eqnarray}
Note that based on estimated   CSI,  $\mu_{m}$ has to be  computed at $R_m$ and  $(\psi_{m}, \rho_m)$ at $R_n$ with $n\neq m$. Therefore, 
using \eqref{Eq:CapSlope}, we propose the following iterative IPC feedback algorithm. 

\noindent {\bf Algorithm $\mathbf{1}$:}
\begin{enumerate}
\item Transmitter $1$ and $2$ arbitrarily select the initial values for $P_1$ and $P_2$, respectively. 

\item The transmitters broadcast their choices of transmission power to the receivers. 

\item Given $(P_1, P_2)$,  the receiver $1$ computes $(\mu_1, \psi_2, \rho_2)$  and feeds back $\mu_1$ and $(\psi_2 - \rho_2)$ to transmitter $1$ and $2$, respectively. Likewise, receiver $2$ computes $(\mu_2, \psi_1,  \rho_1)$ and feeds back     $\mu_2$ and $(\psi_1 - \rho_1)$ to transmitter $2$ and $1$, respectively.

\item Transmitter $1$ and $2$  update $P_1$ and $P_2$, respectively, using \eqref{Eq:CapSlope} and the following equation
\begin{equation}
P_m(k+1)  = \min\l\{\l[P_m(k) + \frac{\partial A(P_1, P_2)}{\partial P_m}\Delta \gamma\r]^+, P_{\max}\r\}\nn
\end{equation}
where $k$ is the iteration index and  $\Delta\gamma$ a  step size.

\item Repeat Steps $2)-4)$ till the maximum number of iterations is performed or the changes on $(P_1, P_2)$ are sufficiently small. 
\end{enumerate}
Note that the IPC-feedback overhead increases linearly with the number of iterations. By choosing an appropriate step size, the convergence of the  above iteration is guaranteed but the converged throughput need not be globally maximum.   

\section{Precoder Feedback Requirements}\label{Section:FbRequire}
In this section, consider the case of $K = N_e$ as in the preceding section  and  the number of  bits  for cooperative precoder feedback is derived under a constraint on the throughput loss due to precoder quantization. 

The  expected precoder  quantization errors is related to the number of feedback bits as  follows. 
Consider the codebook $\mathcal{F}$ of $V$ $L \times N_p$ orthonormal  matrices that is used by each receiver to quantize the inner precoder for the corresponding interferer. Given  $\mathcal{F}$, the quantization error $\epsilon_n$ in \eqref{Eq:QuantErr} is minimized by 
selecting the quantized precoder $\hat{\bF}^{\mathsf{i}}_n$ as
\begin{equation}
\hat{\bF}^{\mathsf{i}}_n =\arg \min_{\bW \in \mathcal{F}} \l(1 - \frac{\|\bW^\dagger\bF_n^{\mathsf{i}}\|^2_{\mathsf{F}}}{N_e}\r), \quad n = 1, 2. 
\end{equation}
The above operation is  equivalent to minimizing the \emph{Chordal distance} between $\hat{\bF}^{\mathsf{i}}_n$ and $\bF^{\mathsf{i}}_n$:  \cite{LovHeaETAL:GrasBeamMultMult:Oct:03}
\begin{equation}\label{Eq:QuantFun}
\hat{\bF}^{\mathsf{i}}_n =\arg \min_{\bW \in \mathcal{F}} d_c(\bW, \bF^{\mathsf{i}}_n) 
\end{equation}
where the chordal distanc $d_c$ is  defined as 
\begin{eqnarray}
d_c(\bW, \bF^{\mathsf{i}}_n)  &=& \frac{1}{\sqrt{2}}\|\bW\bW^\dagger - \bF_n^{\mathsf{i}}(\bF_n^{\mathsf{i}})^\dagger\|_{\mathsf{F}} \nn\\
&=& \sqrt{N_e - \|\bW^\dagger\bF_n^{\mathsf{i}}\|^2_{\mathsf{F}}}. \nn
\end{eqnarray}
The codebook selection in \eqref{Eq:QuantFun} motivates the codebook design based on minimizing the maximum chordal distance between every pair of codebook members \cite{LoveHeath:LimitedFeedbackPrecodSpatialMultiplex:05}. For such a design, the expected quantization error can be upper bounded as \cite{Dai:QuantBoundsGrassmannMIMO}
\begin{equation}\label{Eq:QErr:UB}
\E[\epsilon_n] \leq \frac{\Gamma\l(\frac{1}{Z}\r)}{Z} \beta^{-\frac{1}{Z}}2^{-\frac{B}{Z}} + Le^{-(2^B\beta)^{1-\kappa}}
\end{equation}
where $Z = N_e(L-N_e)$, the number of feedback bits $B = \log_2 V$,   $\beta = \frac{1}{Z!}\prod_{m=1}^{N_e}\frac{(L-m)!}{(N_e-m)!}$,  $\kappa\in (0, 1)$ is a given constant,  and $\Gamma$ denotes the gamma function.

First, we consider a constraint on   the minimum throughput loss due to quantized cooperative precoder feedback, namely 
\begin{equation}
\Delta C = \tilde{C} - \max_{\mathcal{P}}\bar{C}(\mathcal{P})
\end{equation}
with $\bar{C}$ and $\tilde{C}$   given in  \eqref{Eq:SumRate:PerfCSI} and \eqref{Eq:ErgCap:a}, respectively, and $\mathcal{P}$ denotes a power-control policy. To satisfy the constraint $\Delta C \leq c$ with $c> 0$, it is sufficient to equate the following upper bound on $\Delta C$ to $c$:
\begin{equation}
\Delta C \leq \tilde{C} - \bar{C}(P_{\max}, P_{\max})
\end{equation}
where $\bar{C}(P_{\max}, P_{\max})$ corresponds to a sub-optimal power-control policy that fixes the power of both transmitters  at the maximum. 
The above upper bound has a similar form as the throughput loss for multi-antenna downlink with limited feedback as defined in \cite{RavJindal:LimFbDiagonalMIMOBC:2008}. Thus, the following result can be proved following a similar procedure as \cite[Theorem~2]{RavJindal:LimFbDiagonalMIMOBC:2008}. 
\begin{corollary} \label{Cor:B:CapMaxPwr} For large $P_{\max}$, choosing  the  number of bits for cooperative precoder feedback as  
\begin{equation}
B = Z\log_2(\nu P_{\max})  - Z\log_2\l(2^{\frac{c}{2N_e}}-1\r)+ \omega \label{Eq:B:OptimCap}
\end{equation}
ensures that 
\begin{equation}
\Delta C \leq c + o(1), \quad P_{\max}\rightarrow \infty\label{Eq:CapLoss:UB}
\end{equation}
where $\omega = Z\log_2\frac{N_e\Gamma\l(\frac{1}{Z}\r)\E[\check{\lambda}_{K-N_e+1}]}{Z\beta^{\frac{1}{Z}}}$. 
\end{corollary}A few remarks are in order: 
\begin{enumerate}
\item For large $P_{\max}$,  $B\approx Z\log_2 P_{\max} $. For small $P_{\max}$, the network is noise limited and the number of precoder feedback bits  can be kept small. 
\item For the case of fixed interference margin,  it can be also proved that $B\approx Z\log_2 P_{\max} $ for large $P_{\max}$ following a similar procedure as Corollary~\ref{Cor:B:CapMaxPwr}. 

\item The upper bound on the capacity loss approaches  $c$ as  $P_{\max}$ increases.  

\item { The  feedback-bit scaling obtained in \cite{RavJindal:LimFbDiagonalMIMOBC:2008} for the MIMO downlink system is similar to that in  \eqref{Eq:B:OptimCap} despite the difference in system configuration. Specifically, it is shown in \cite[Theorem~2]{RavJindal:LimFbDiagonalMIMOBC:2008} that the number of precoder-feedback bits  per user should scale as $B\approx \tilde{N}(\tilde{M}-\tilde{N})\log \tilde{P}$ for large $\tilde{P}$   so as to constrain the sum-throughput loss,  where $\tilde{M}$ and $\tilde{N}$ are the numbers of antennas at the base station and each mobile, respectively, and $\tilde{P}$ is the total transmission power at the base station. The above  similarity  rises from the fact that both the proposed  precoding and the block-diagonalization precoding in  \cite{RavJindal:LimFbDiagonalMIMOBC:2008} are designed using the zero-forcing criterion to null multiuser interference. }

\end{enumerate}

\begin{figure}
\begin{center}
\includegraphics[width=12cm]{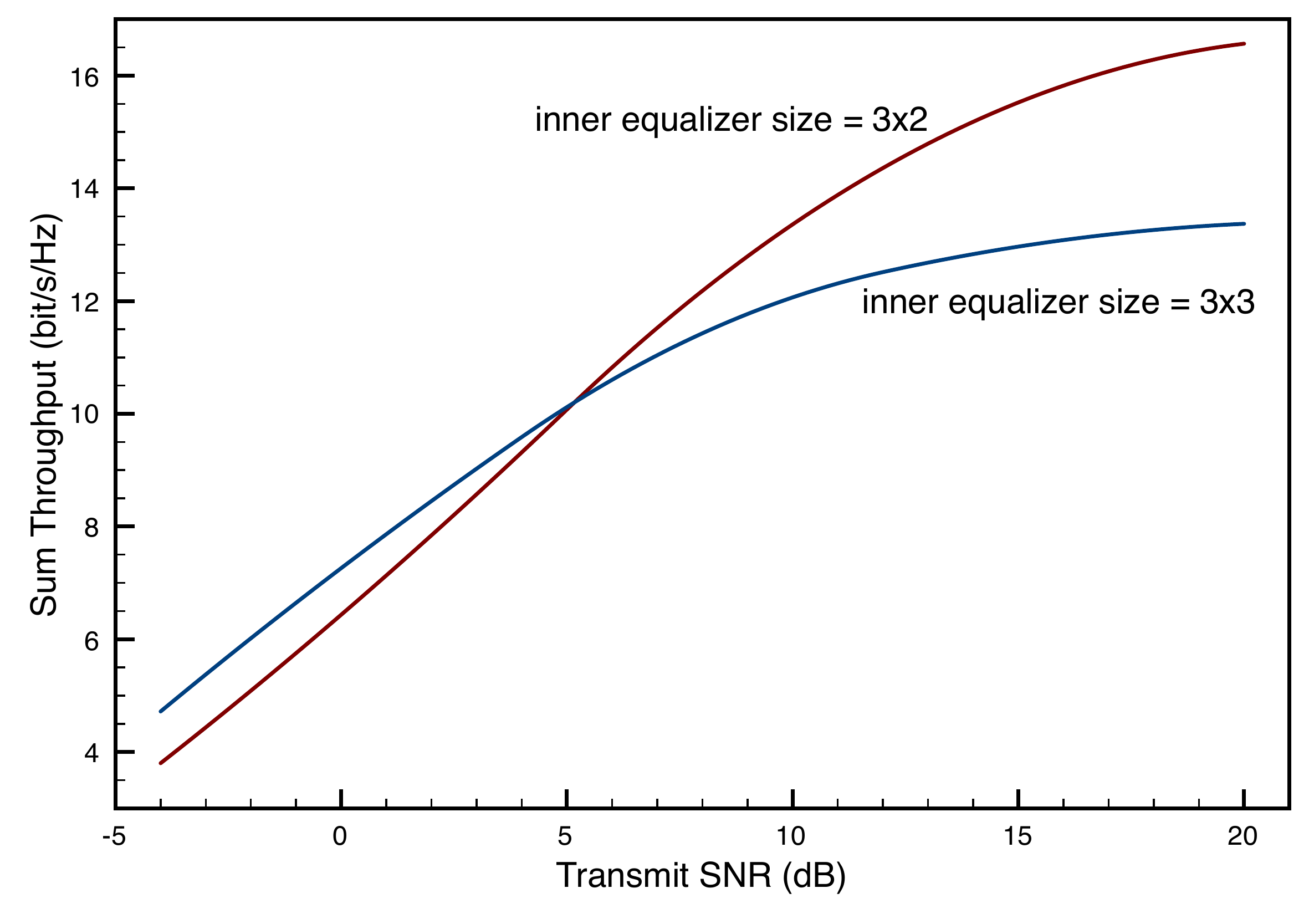}
\caption{Effect of the width  of inner equalizer matrices, $N_e$,  on the sum throughput for different  SNRs.  The  transmission power is fixed at  $P_{\max}$.   }\label{Fig:NEffect}\end{center}
\end{figure}

\section{Simulation Results}\label{Section:Simulation}
In the simulation, the codebook for quantizing the feedback precoders is randomly generated as in \cite{YeungLove:RandomVQBeamf:05} and the system performance is averaged over a larger number of codebook realizations. The simulation parameters are set as follows unless specified otherwise.  The numbers of antennas at each transmitter and receiver are $L=6$ and  $K = 3$, respectively.  The number of data stream per user is $M=2$. The size of inner precoder is fixed as $6\times 3$. The path-loss factor $\nu$ is set as $\nu=0.5$.   Power control based on interference margin uses $\tau = 2$. The   number of cooperative-feedback bit is   $B=8$.

\subsection{Performance Evaluation}

The size of inner equalizer  determines the allocation of DoF at each receiver for mitigating residual interference and for enhancing array gain. The sum throughput for two different inner-equalizer sizes, namely $3\times 2$ and $3\times 3$, is  compared in Fig.~\ref{Fig:NEffect} with transmission power of each transmitter fixed as $P_{\max}$. The larger inner precoder ($3\times 3$) allocates more DoF for achieving array gain and it can be observed to  increase the  throughput for low SNRs where noise dominates residual interference. However, the  smaller inner precoder is preferred for high SNRs since it allows more DoF for mitigating residual interference. 

\begin{figure}[t]
\begin{center}
\includegraphics[width=12cm]{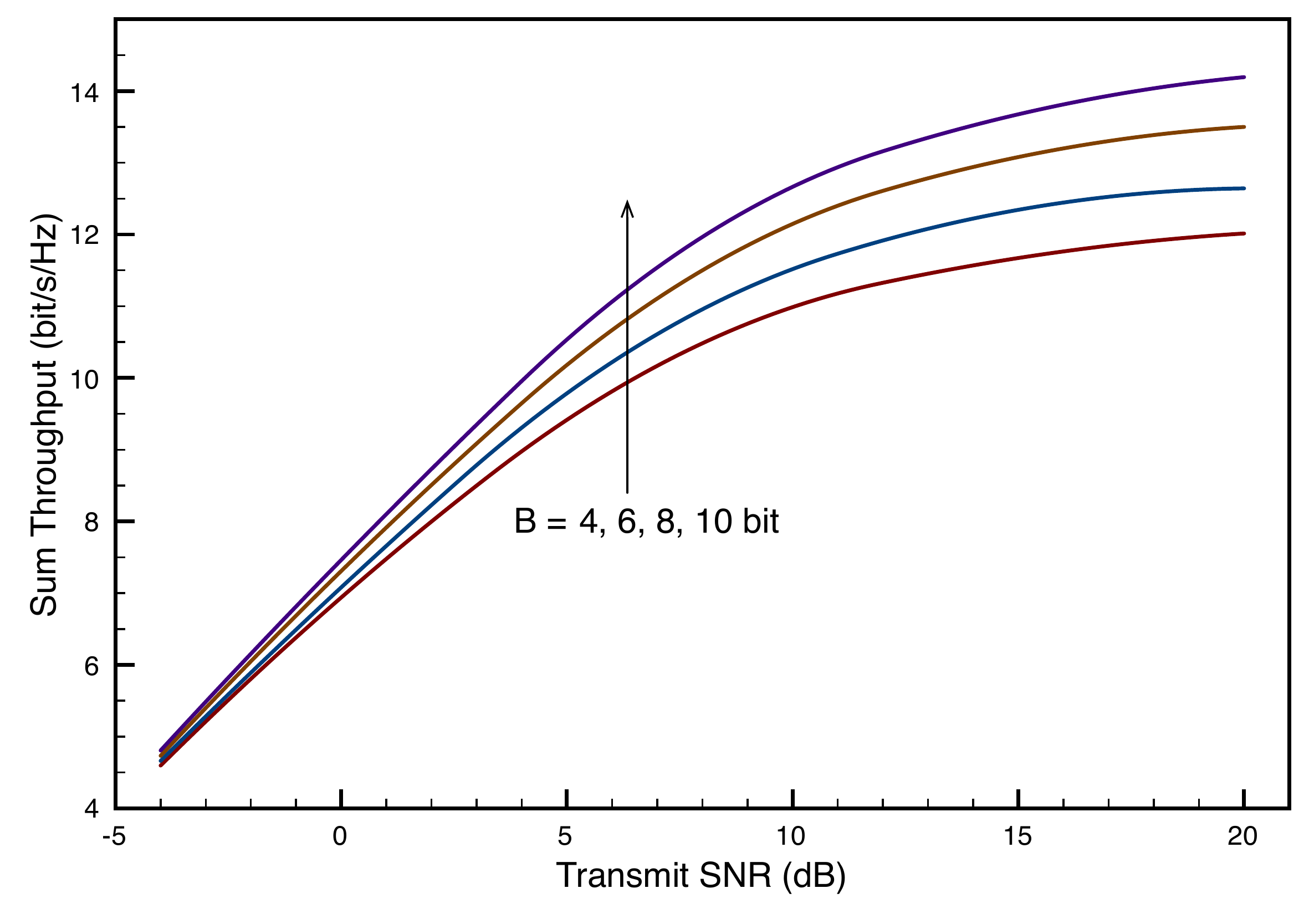}
\caption{{ Comparison of achievable sum throughput of the proposed  precoding   design for  different numbers of cooperative-feedback bit $B$. The  transmission power is fixed as $P_{\max}$ and the inner equalizer  has the size of $N_e\times K = 3\times 3$.}  }
\label{Fig:Cap:Var:B}
\end{center}
\end{figure}

Fig.~\ref{Fig:Cap:Var:B} compares the sum  throughput of the proposed transceiver design for the  cooperative-feedback bit $B=\{4, 6, 8, 10\}$. The transmission power of each transmitter is fixed as $P_{\max}$ and the inner equalizer  has the size of $N_e\times K = 3\times 3$.  It can be observed that the increment of every $2$ cooperative-feedback bits increases the sum throughput by about  $0.7$ bit/s/Hz for high SNRs. 

\begin{figure}
\centering
\includegraphics[width=12cm]{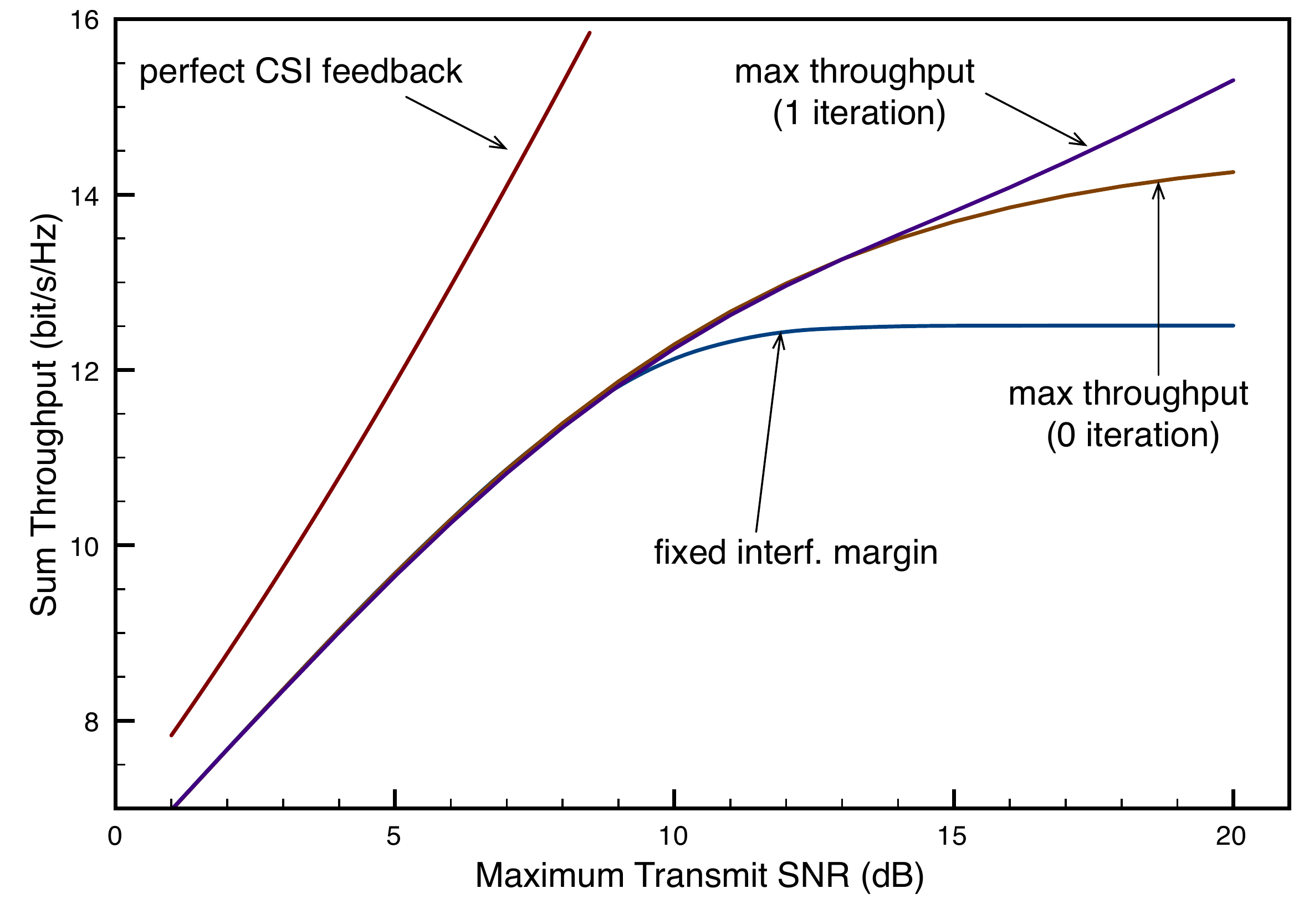}
  \caption{Comparison of achievable sum throughput between different IPC feedback algorithms. The inner equalizer has the size of $N_e\times K = 3\times 3$. For IPC feedback Algorithm~$1$ that maximizes achievable throughput,   the step size $\Delta\gamma$ for updating transmission power is  $P_{\max}$; the initial  transmission power is set as $P_{\max}/2$.    }\label{Fig:IPC:SumCap}
\end{figure}

Fig.~\ref{Fig:IPC:SumCap} compares the sum  throughput of different IPC feedback algorithms where the inner equalizer  size is set as  $ 3\times 3$. Residual interference between links is observed to decrease the  throughput dramatically with respect to perfect CSI feedback.  For large $P_{\max}$, the  IPC feedback Algorithm~$1$ designed for maximizing the achievable throughput is observed to provide substantial  throughput gain over that  based on fixed interference margin. Moreover,  iterations for the  IPC  feedback Algorithm~$1$ are observed to give significant gain only at high SNRs. 

\subsection{Comparison with a Conventional Transceiver Design}

\begin{figure}[t]
\begin{center}
\includegraphics[width=12cm]{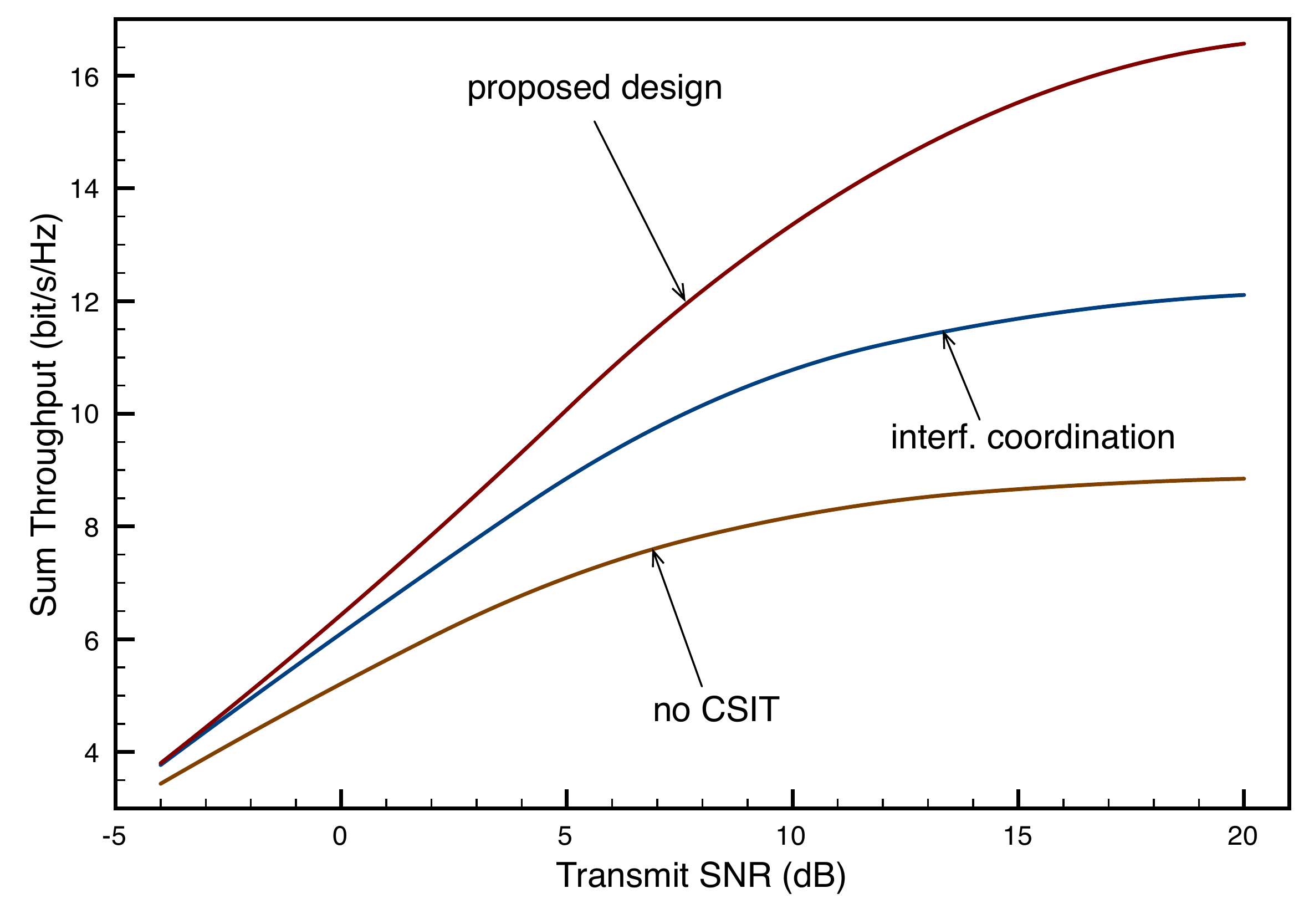}
\caption{{ Comparison of  sum throughput between the proposed  and two conventional precoding algorithms for the MIMO interference channel, namely interference coordination and the case of no CSIT. The  transmission power is fixed at  $P_{\max}$. Moreover, for the proposed deign, the inner equalizer size is set as $3\times 3$. }  }
\label{Fig:Cap:Cmp}
\end{center}
\end{figure}

The proposed precoding algorithm is compared with the conventional \emph{interference coordination} \cite{Gesbert:MultiCellMIMOCooperativeNetworks:2010} in terms of sum throughput with quantized  cooperative feedback. The interference coordination algorithm attempts to align the interference by precoding such that at each receiver interference is observed only over the last $M$ antennas and the signals received over the first $(L-M)$ antennas are free of interference. To this end,  the $L\times M$ precoder of user $m$,  denoted as $\bF'_m$, is chosen to be orthogonal to the channel sub-matrix $[\bH_{nm}]_{1:(L-M)}^{\text{row}}$ with $n\neq m$ where $[\bX]_{m:n}^{\text{row}}$ denotes a sub-matrix comprising row $n$ to $m$ of a matrix  $\bX$; the quantized version $\hat{\bF}'_m$ of $\bF'_m$ is fed back from receiver $n$ to  transmitter $m$ for precoding. Furthermore, the receiver $\bG_m'$ of user $m$  decouples the data streams by zero-forcing: 
\begin{equation}\label{Eq:Algo:a}
\bG_m' = \l([\bH_{mm}]_{1:(L-M)}^{\text{row}} \hat{\bF}_m'\r)\l[\l([\bH_{mm}]_{1:(L-M)}^{\text{row}} \hat{\bF}'_m\r)^\dagger\l([\bH_{mm}]_{1:(L-M)}^{\text{row}} \hat{\bF}_m'\r)\r]^{-1}.  
\end{equation}
Also considered in the comparison is the case of no CSIT where the precoder $\{\hat{\bF}'_m\}$ are arbitrarily chosen to be independent with all channels.  The achievable sum throughput for the proposed and  conventional algorithms are compared in Fig.~\ref{Fig:Cap:Cmp}. The transmission power of each transmitter is $P_{\max}$; for the proposed deign, the inner equalizer size is set as $3\times 3$. It can be observed from Fig.~\ref{Fig:Cap:Cmp} that the proposed cooperative-feedback design yields dramatic throughput gains over the conventional algorithms. In particular, the gain over interference coordination is as large as about $4$ bit/s/Hz for high SNRs.  The performance gains of the proposed design result from the joint tuning of precoders and equalizers for simultaneously suppressing residual interference and harvesting diversity gain, which is enabled by the proposed inner/outer transceiver structure.

\section{Conclusion}\label{Section:Conclusion}
We have proposed a systematic  design of linear precoders and equalizers for the two-user MIMO interference channel with finite-rate cooperative precoder feedback. This design suppresses  residual interference due to feedback precoder quantization. Building upon  the above design, we have further proposed  scalar cooperative feedback algorithms for controlling transmission power based on  different criteria including fixed interference margin and  maximum sum throughput. Finally, we have derived the scaling of the number of cooperative precoder-feedback bits
under a the constraint on the sum throughput loss. Possible extensions of the current work include generalizing    the proposed algorithms to the interference channel with an  arbitrary number of users and relaxing the current zero-forcing criterion on the precoder design.

\appendix

Lemma~\ref{Lemma:SumCap:IM} is proved  as follows. 
For convenience, the achievable throughput in  \eqref{Eq:SumCap:LB} can be written as 
\begin{equation}
A_{\mathsf{IM}} = \sum_{m=1}^2 \sum_{\ell = 1}^M A_{m}^{[\ell]} \label{Eq:SumCap:LB:a}
\end{equation}
where 
\begin{equation}
A_{m}^{[\ell]} = \E\l[\log_2\l(1 + \frac{\min(\eta_m, P_{\max}) \lambda_{mm}^{[\ell]}}{1+\tau}\r)\r]. 
\end{equation}
Expand $A_{m}^{[\ell]}$ as
\begin{equation}\begin{aligned}
&&A_{m}^{[\ell]} =&\ \E\l[\log_2\l(1 + \frac{\eta_m \lambda_{mm}^{[\ell]}}{1+\tau}\r)\mid \eta_m\leq P_{\max}\r]\Pr(\eta_m\leq P_{\max}) + \\
&&&\ \E\l[\log_2\l(1 + \frac{P_{\max} \lambda_{mm}^{[\ell]}}{1+\tau}\r)\r]\Pr(\eta_m> P_{\max}). 
\end{aligned}\label{Eq:CapComp}
\end{equation}
Using \eqref{Eq:IPC}, the first term $\Phi_1$ of $A_{m}^{[\ell]}$ in \eqref{Eq:CapComp} can be rewritten as 
\begin{equation}
\Phi_1 = \E\l[\log_2\l(1 + \frac{\tau \acute{\lambda}_{\ell}}{(1+\tau)N_p\nu\check{\lambda}_{K-N_e+1}\epsilon_m}\r)\r] - \Phi_3 \label{Eq:CapComp:Term1:a}
\end{equation}
with  $\Phi_3$ defined and simplified as 
\begin{eqnarray}
\Phi_3 &=& \E\l[\int_0^{\frac{\tau}{N_p\nu\epsilon_mP_{\max}}}\log_2\l(1 + \frac{\tau \acute{\lambda}_{\ell}}{(1+\tau)N_p\nu\epsilon_mx}\r)f_{\check{\lambda}_{K-N_e+1}}(x) dx\r]\nn\\
&=& \E\l\{\int_0^{\frac{\tau}{N_p\nu\epsilon_mP_{\max}}}\l[O(1) + \log_2\frac{1}{x}\r]f_{\check{\lambda}_{K-N_e+1}}(x) dx\r\},\quad P_{\max}\rightarrow\infty\nn\\
&\overset{(a)}{=}& \E\l\{\int_0^{\frac{\tau}{N_p\nu\epsilon_mP_{\max}}}o(1)\times [\check{a}_{K-N_e+1}x^{N_e^2-2} + o(x^{N_e^2-2})] dx\r\}\nn\\
&=& o\l(P_{\max}^{-N_e^2+1}\r)\label{Eq:CapComp:Term3}
\end{eqnarray}
where $(a)$ is obtained using Lemma~\ref{Lem:Wishart} and $x\log_2\frac{1}{x} = o(1)$ for $x\rightarrow 0 $. Similarly, we obtain the second  term $\Phi_2$ of $A_{m}^{[\ell]}$ in \eqref{Eq:CapComp} as
\begin{equation}
\Phi_2 = o\l(P_{\max}^{-N_e^2+1}\r),\quad P_{\max}\rightarrow \infty. \label{Eq:CapComp:Term2:a}
\end{equation}
Substituting \eqref{Eq:CapComp:Term1:a},   \eqref{Eq:CapComp:Term3}, and \eqref{Eq:CapComp:Term2:a} into \eqref{Eq:CapComp} and then \eqref{Eq:SumCap:LB:a} yields  \eqref{Eq:Thput:IM:LargeP}, completing  the proof. 

\bibliographystyle{ieeetr}

\begin{biography}[{\includegraphics[width=1in, clip, keepaspectratio]{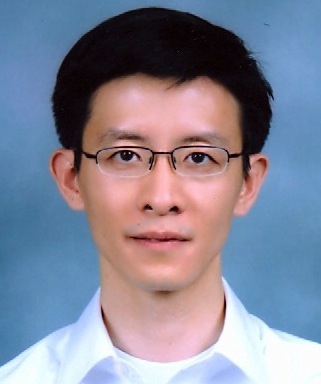}}]{Kaibin Huang}  (S'05--M'08) received the B.Eng. (first-class hons.) and the M.Eng. from the National University of Singapore in 1998 and 2000, respectively, and the Ph.D. degree from The University of Texas at Austin (UT Austin) in 2008, all in electrical engineering.

Since Mar. 2009, he has been an assistant professor in the School of Electrical and Electronic Engineering at Yonsei University, Seoul, Korea. From Jun. 2008 to Feb. 2009, he was a Postdoctoral Research Fellow in the Department of Electrical and Computer Engineering at the Hong Kong University of Science and Technology. From Nov. 1999 to Jul. 2004, he was an Associate Scientist at the Institute for Infocomm Research in Singapore. He frequently serves on the technical program committees of major IEEE conferences in wireless communications. Recently, he is the technical co-chair of IEEE CTW 2013 and the track chairs of IEEE Asilomar 2011 and IEEE WCNC 2011. He is an editor for the IEEE Wireless Communications Letters and also Journal of Communication and Networks. Dr. Huang received the Outstanding Teaching Award from Yonsei, Motorola Partnerships in Research Grant, the University Continuing Fellowship at UT Austin, and the Best Paper award at IEEE GLOBECOM 2006. His research interests focus on multi-antenna limited feedback techniques and the analysis and design of wireless networks using stochastic geometry.
\end{biography}
\vfill 

\begin{biography}[{\includegraphics[width=1in, clip, keepaspectratio]{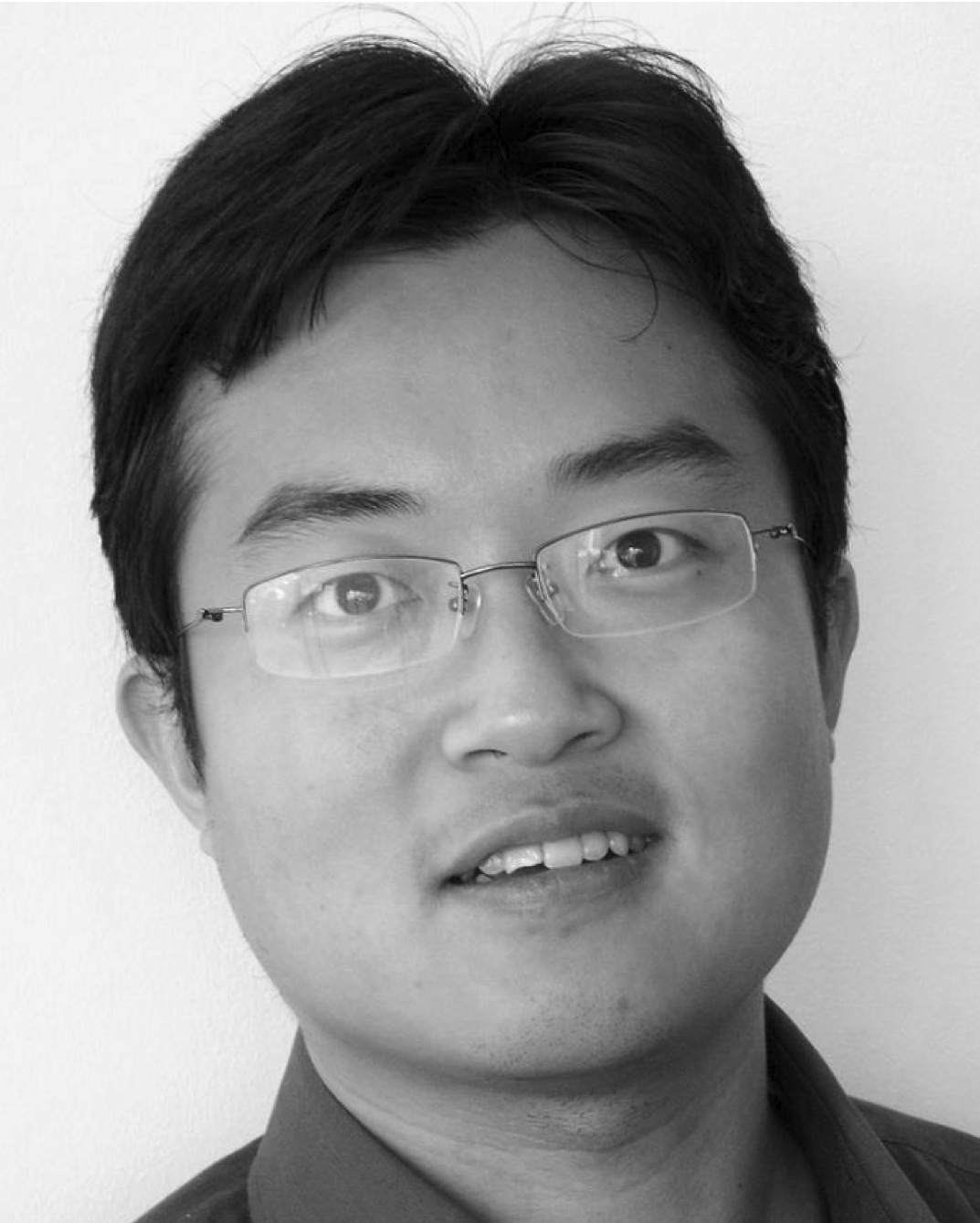}}]{Rui Zhang}  (S'00--M'07) received the B.Eng. (First-Class Hons.) and M.Eng. degrees from the National University of Singapore in 2000 and 2001, respectively, and the Ph.D. degree from the Stanford University, Stanford, CA USA, in 2007, all in electrical engineering.

Since 2007, he has worked with the Institute for Infocomm Research, A*STAR, Singapore, where he is now a Senior Research Scientist. Since 2010, he has also held an Assistant Professorship position with the Department of Electrical and Computer Engineering at the National University of Singapore. He has authored/coauthored over 100 internationally refereed journal and conference papers. His current research interests include wireless communications (e.g., multiuser MIMO, cognitive radio, cooperative communication, energy efficiency and energy harvesting), wireless power and information transfer, smart grid, and optimization theory for applications in communication and power networks.

Dr. Zhang was the co-recipient of the Best Paper Award from the IEEE PIMRC (2005). He was Guest Editors of the EURASIP Journal on Applied Signal Processing Special Issue on Advanced Signal Processing for Cognitive Radio Networks (2010), the Journal of Communications and Networks (JCN) Special Issue on Energy Harvesting in Wireless Networks (2011), and the EURASIP Journal on Wireless Communications and Networking Special Issue on Recent Advances in Optimization Techniques in Wireless Communication Networks (2012). He has also served for various IEEE conferences as Technical Program Committee (TPC) members and Organizing Committee members. He was the recipient of the 6th IEEE ComSoc Asia-Pacific Best Young Researcher Award (2010), and the Young Investigator Award of National University of Singapore (2011). He is an elected member for IEEE Signal Processing Society SPCOM Technical Committee.

\end{biography}

\vfill 

\end{document}

%% file: header.tex
\newtheorem{lemma}{Lemma}
\newtheorem{corollary}{Corollary}

\def\E{\mathsf{E}}

\def\phi{\varphi}

\def\SINR{\mathsf{SINR}}
\def\SNR{\mathsf{SNR}}

\def\l{\left}
\def\r{\right}
\def\({\left(}
\def\){\right)}

\def\b0{{\mathbf{0}}}

\def\bA{{\mathbf{A}}}
\def\bB{{\mathbf{B}}}
\def\bC{{\mathbf{C}}}

\def\bF{{\mathbf{F}}}
\def\bG{{\mathbf{G}}}
\def\bH{{\mathbf{H}}}
\def\bI{{\mathbf{I}}}

\def\bU{{\mathbf{U}}}
\def\bV{{\mathbf{V}}}
\def\bW{{\mathbf{W}}}
\def\bX{{\mathbf{X}}}

\newcommand{\nn}{\nonumber}

%% file: iChannel_Feedback_ArXiv.bbl
\begin{thebibliography}{10}

\bibitem{CadJafar:InterfAlignment:2007}
V.~R. Cadambe and S.~A. Jafar, ``Interference alignment and the degrees of
  freedom for the {K} user interference channel,'' {\em {IEEE} Trans. on
  Information Theory}, vol.~54, pp.~3425--3441, Aug. 2008.

\bibitem{ZakhourGesbert:DistMutlicellMISOPrecodingLayerVirtualSINR}
R.~Zakhour and D.~Gesbert, ``Distributed multicell-{MISO} precoding using the
  layered virtual {SINR} framework,'' {\em IEEE Trans. Wireless Comm.}, vol.~9,
  pp.~2444--2448, Aug. 2010.

\bibitem{DahYu:CoordBeamformMulticell:2010}
H.~Dahrouj and W.~Yu, ``Coordinated beamforming for the multicell multi-antenna
  wireless system,'' {\em IEEE Trans. on Wireless Comm.}, vol.~9,
  pp.~1748--1759, May 2010.

\bibitem{ChaeHeath:InterfAwareCoordBeamformTwoCell}
C.~B. Chae, I.~Hwang, R.~W. Heath~Jr., and V.~Tarokh, ``Interference
  aware-coordinated beamforming system in a two-cell environment,'' {\em
  Technical Report (Available:
  http://nrs.harvard.edu/urn-3:HUL.InstRepos:3293263)}.

\bibitem{ZhangCui:CoopIMMISOBeamform}
R.~Zhang and S.~Cui, ``Cooperative interference management with {MISO}
  beamforming,'' {\em IEEE Trans. on Sig. Proc.}, vol.~58, pp.~5450--5458, Oct.
  2010.

\bibitem{Love:OverviewLimitFbWirelssComm:2008}
D.~J. Love, R.~W. Heath~Jr., V.~K.~N. Lau, D.~Gesbert, B.~D. Rao, and
  M.~Andrews, ``An overview of limited feedback in wireless communication
  systems,'' {\em IEEE Journal on Selected Areas in Comm.}, vol.~26, no.~8,
  pp.~1341--1365, 2008.

\bibitem{LovHeaETAL:GrasBeamMultMult:Oct:03}
D.~J. Love, R.~W. Heath~Jr., and T.~Strohmer, ``Grassmannian beamforming for
  {MIMO} wireless systems,'' {\em IEEE Trans. on Information Theory}, vol.~49,
  pp.~2735--2747, Oct. 2003.

\bibitem{Lau:MIMOBlockFadingFbLinkCapConst:04}
V.~K.~N. Lau, Y.~Liu, and T.-A. Chen, ``On the design of {MIMO} block-fading
  channels with feedback-link capacity constraint,'' {\em IEEE Trans. on
  Comm.}, vol.~52, pp.~62--70, Jan. 2004.

\bibitem{LoveHeath:LimitedFeedbackPrecodSpatialMultiplex:05}
D.~J. Love and R.~W. Heath~Jr., ``Limited feedback unitary precoding for
  spatial multiplexing systems,'' {\em IEEE Trans. on Information Theory},
  vol.~51, pp.~1967--1976, Aug. 2005.

\bibitem{Gesbert:ShiftMIMOParadigm:2007}
D.~Gesbert, M.~Kountouris, R.~W. Heath~Jr., C.-B. Chae, and T.~Salzer, ``From
  single user to multiuser communications: Shifting the {MIMO} paradigm,'' {\em
  IEEE Signal Proc. Magazine}, vol.~24, no.~5, pp.~36--46, 2007.

\bibitem{Jindal:MIMOBroadcastFiniteRateFeedback:06}
N.~Jindal, ``{MIMO} broadcast channels with finite-rate feedback,'' {\em IEEE
  Trans. on Information Theory}, vol.~52, pp.~5045--5060, Nov. 2006.

\bibitem{SharifHassibi:CapMIMOBroadcastPartSideInfo:Feb:05}
M.~Sharif and B.~Hassibi, ``On the capacity of {MIMO} broadcast channels with
  partial side information,'' {\em IEEE Trans. on Information Theory}, vol.~51,
  pp.~506--522, Feb. 2005.

\bibitem{Huang:OrthBeamSDMALimtFb:07}
K.~Huang, J.~G. Andrews, and R.~W. Heath~Jr., ``Performance of orthogonal
  beamforming for {SDMA} systems with limited feedback,'' {\em IEEE Trans. on
  Veh. Technology}, vol.~58, pp.~152--164, Jan. 2009.

\bibitem{HuangZhang:CoopFeedbackCognitiveRadio}
K.~Huang and R.~Zhang, ``Cooperative feedback for multi-antenna cognitive radio
  networks,'' {\em IEEE Trans. on Sig. Proc.}, vol.~59, pp.~747--758, Feb.
  2011.

\bibitem{Schmit:DistResourceAllocationPrice:2009}
D.~A. Schmidt, C.~Shi, R.~A. Berry, M.~L. Honig, and W.~Utschick, ``Distributed
  resource allocation schemes: Pricing algorithms for power control and
  beamformer design in interference networks,'' {\em {IEEE} Signal Proc.
  Magazine}, vol.~26, pp.~53--63, May 2009.

\bibitem{Jorswieck:CompleteCharactParetoMISOInterf:2008}
E.~A. Jorswieck, E.~G. Larsson, and D.~Danev, ``Complete characterization of
  the pareto boundary for the miso interference channel,'' {\em {IEEE} Tarns.
  on Signal Proc.}, vol.~56, pp.~5292--5296, Oct. 2008.

\bibitem{Gesbert:MultiCellMIMOCooperativeNetworks:2010}
D.~Gesbert, S.~Hanly, H.~Huang, S.~Shitz, O.~Simeone, and W.~Yu, ``Multi-cell
  {MIMO} cooperative networks: A new look at interference,'' {\em {IEEE}
  Journal on Sel. Areas in Comm.}, vol.~28, pp.~1380--1408, Sep. 2010.

\bibitem{Chae:BDVPMuMIMO:2008}
C.~Chae, S.~Shim, and R.~Heath, ``Block diagonalized vector perturbation for
  multiuser {MIMO} systems,'' {\em {IEEE} Trans. on Wireless Comm.}, vol.~7,
  no.~11, pp.~4051--4057, 2008.

\bibitem{RavJindal:LimFbDiagonalMIMOBC:2008}
N.~Ravindran and N.~Jindal, ``Limited feedback-based block diagonalization for
  the {MIMO} broadcast channel,'' {\em IEEE Journal on Sel. Areas in Comm.},
  vol.~26, pp.~1473--1482, Aug. 2008.

\bibitem{PaulrajBook}
A.~Paulraj, R.~Nabar, and D.~Gore, {\em Introduction to Space-Time Wireless
  Communications}.
\newblock Cambridge University Press, 2003.

\bibitem{OrdeonezPalomar:HighSNRPerformMIMO:2007}
L.~G. Ord{\'o}{\~n}ez, D.~P. Palomar, A.~Pag{\`e}s-Zamora, and J.~R. Fonollosa,
  ``High-{SNR} analytical performance of spatial multiplexing {MIMO} systems
  with {CSI},'' {\em IEEE Trans. on Signal Proc.}, vol.~55, pp.~5447--5463,
  Nov. 2007.

\bibitem{GjenGesbert:BinaryPowerConrolInterferenceLinks:2008}
A.~Gjendemsjoe, D.~Gesbert, G.~Oien, and S.~Kiani, ``Binary power binary power
  control for sum rate maximization over multiple interfering links,'' {\em
  IEEE Trans. on Wireless Comm.}, vol.~7, pp.~3164--3173, Aug. 2008.

\bibitem{Foschini:AutoPowerControl:1993}
G.~Foschini and Z.~Miljanic, ``A simple distributed autonomous power control
  algorithm and its convergence,'' {\em IEEE Trans. on Veh. Technology},
  vol.~42, pp.~641--646, Apr. 1993.

\bibitem{Dai:QuantBoundsGrassmannMIMO}
W.~Dai, Y.~Liu, and B.~Rider, ``Quantization bounds on {Grassmann} manifolds
  and applications to {MIMO} systems,'' {\em {IEEE} Trans. on Information
  Theory}, vol.~54, pp.~1108--1123, Mar. 2008.

\bibitem{ChoiHeath:InterpTxBeamMIMOOFDMLimFb:2005}
J.~Choi and R.~W. Heath~Jr., ``Interpolation based transmit beamforming for
  {MIMO}-{OFDM} with limited feedback,'' {\em {IEEE} Tarns. on Signal Proc.},
  vol.~53, pp.~4125--35, Nov. 2005.

\bibitem{Huang:LimFbBeamformTemporallyCorrChan:2009}
K.~Huang, R.~W. Heath~Jr., and J.~G. Andrews, ``Limited feedback beamforming
  over temporally-correlated channels,'' {\em {IEEE} Tarns. on Signal Proc.},
  vol.~57, pp.~1959--1975, May 2009.

\bibitem{HuangLau:EventDrivenFeedbackControlBeamforming}
K.~Huang, V.~K.~N. Lau, and D.~Kim, ``Event-driven optimal feedback control for
  multi-antenna beamforming,'' {\em IEEE Trans. on Signal Proc.}, vol.~58,
  pp.~3298--3312, Jun. 2010.

\bibitem{YeungLove:RandomVQBeamf:05}
C.~K. Au-Yeung and D.~J. Love, ``On the performance of random vector
  quantization limited feedback beamforming in a {MISO} system,'' {\em IEEE
  Trans. on Wireless Comm.}, vol.~6, pp.~458--462, Feb. 2007.

\end{thebibliography}
